\documentclass[11pt,fleqn]{article}
\usepackage{amsfonts, epsfig, amssymb, amsmath}
\usepackage[normalem]{ulem}
\usepackage{color}
\newcommand{\ol}{\setlength{\itemsep}{0pt.}\begin{enumerate}}
\newcommand{\eol}{\end{enumerate}\setlength{\itemsep}{-\parsep}}
\newcommand{\ignore}[1]{}
\title{On the difficulty to beat the first linear programming bound for binary codes}
\author{Alex Samorodnitsky}

\begin{document}
\date{}
\maketitle

%  THEOREM-LIKE ENVIRONMENTS

\newtheorem{THEOREM}{Theorem}[section]
\newenvironment{theorem}{\begin{THEOREM} \hspace{-.85em} {\bf :}
}%
                        {\end{THEOREM}}
\newtheorem{LEMMA}[THEOREM]{Lemma}
\newenvironment{lemma}{\begin{LEMMA} \hspace{-.85em} {\bf :} }%
                      {\end{LEMMA}}
\newtheorem{COROLLARY}[THEOREM]{Corollary}
\newenvironment{corollary}{\begin{COROLLARY} \hspace{-.85em} {\bf
:} }%
                          {\end{COROLLARY}}
\newtheorem{PROPOSITION}[THEOREM]{Proposition}
\newenvironment{proposition}{\begin{PROPOSITION} \hspace{-.85em}
{\bf :} }%
                            {\end{PROPOSITION}}
\newtheorem{DEFINITION}[THEOREM]{Definition}
\newenvironment{definition}{\begin{DEFINITION} \hspace{-.85em} {\bf
:} \rm}%
                            {\end{DEFINITION}}
\newtheorem{EXAMPLE}[THEOREM]{Example}
\newenvironment{example}{\begin{EXAMPLE} \hspace{-.85em} {\bf :}
\rm}%
                            {\end{EXAMPLE}}
\newtheorem{CONJECTURE}[THEOREM]{Conjecture}
\newenvironment{conjecture}{\begin{CONJECTURE} \hspace{-.85em}
{\bf :} \rm}%
                            {\end{CONJECTURE}}
\newtheorem{MAINCONJECTURE}[THEOREM]{Main Conjecture}
\newenvironment{mainconjecture}{\begin{MAINCONJECTURE} \hspace{-.85em}
{\bf :} \rm}%
                            {\end{MAINCONJECTURE}}
\newtheorem{PROBLEM}[THEOREM]{Problem}
\newenvironment{problem}{\begin{PROBLEM} \hspace{-.85em} {\bf :}
\rm}%
                            {\end{PROBLEM}}
\newtheorem{QUESTION}[THEOREM]{Question}
\newenvironment{question}{\begin{QUESTION} \hspace{-.85em} {\bf :}
\rm}%
                            {\end{QUESTION}}
\newtheorem{REMARK}[THEOREM]{Remark}
\newenvironment{remark}{\begin{REMARK} \hspace{-.85em} {\bf :}
\rm}%
                            {\end{REMARK}}
%\newenvironment{proof}{\noindent {\bf Proof:} \hspace{.677em}}%
%                      {}

%theorem
\newcommand{\thm}{\begin{theorem}}
%lemma
\newcommand{\lem}{\begin{lemma}}
%proposition
\newcommand{\pro}{\begin{proposition}}
%definition
\newcommand{\dfn}{\begin{definition}}
%remark
\newcommand{\rem}{\begin{remark}}
%example
\newcommand{\xam}{\begin{example}}
%conjecture
\newcommand{\cnj}{\begin{conjecture}}
%main_conjecture
\newcommand{\mcnj}{\begin{mainconjecture}}
%problem
\newcommand{\prb}{\begin{problem}}
%question
\newcommand{\que}{\begin{question}}
%corollary
\newcommand{\cor}{\begin{corollary}}
%proof
\newcommand{\prf}{\noindent{\bf Proof:} }
%end theorem
\newcommand{\ethm}{\end{theorem}}
%end lemma
\newcommand{\elem}{\end{lemma}}
%end proposition
\newcommand{\epro}{\end{proposition}}
%end definition
\newcommand{\edfn}{\bbox\end{definition}}
%end remark
\newcommand{\erem}{\bbox\end{remark}}
%end example
\newcommand{\exam}{\bbox\end{example}}
%end conjecture
\newcommand{\ecnj}{\bbox\end{conjecture}}
%end main_conjecture
\newcommand{\emcnj}{\bbox\end{mainconjecture}}
%end problem
\newcommand{\eprb}{\bbox\end{problem}}
%end question
\newcommand{\eque}{\bbox\end{question}}
%end corollary
\newcommand{\ecor}{\end{corollary}}
%end proof
\newcommand{\eprf}{\bbox}
%begin equation
\newcommand{\beqn}{\begin{equation}}
%end equation
\newcommand{\eeqn}{\end{equation}}
% white box
\newcommand{\wbox}{\mbox{$\sqcap$\llap{$\sqcup$}}}
%black box
\newcommand{\bbox}{\vrule height7pt width4pt depth1pt}
\newcommand{\qed}{\bbox}
% \sup will be used for superscript.
\def\sup{^}

\def\H{\{0,1\}^n}

\def\S{S(n,w)}

\def\g{g_{\ast}}
\def\xop{x_{\ast}}
\def\y{y_{\ast}}
\def\z{z_{\ast}}

\def\f{\tilde f}

\def\n{\lfloor \frac n2 \rfloor}

\def \E{\mathop{{}\mathbb E}}
\def \R{\mathbb R}
\def \Z{\mathbb Z}
\def \F{\mathbb F}
\def \T{\mathbb T}

\def \x{\textcolor{red}{x}}
\def \r{\textcolor{red}{r}}
\def \Rc{\textcolor{red}{R}}

\def \noi{{\noindent}}

\def \iff{~~~~\Longleftrightarrow~~~~}

\def\myblt{\noi --\, }

\def \queq {\quad = \quad}

\def\<{\left<}
\def\>{\right>}
\def \({\left(}
\def \){\right)}

\def \e{\epsilon}
\def \l{\lambda}

\def\myblt{\noi --\, }

% Defining Tchebyshef polynomial
\def\Tp{Tchebyshef polynomial}
\def\Tps{TchebysDeto be the maximafine $A(n,d)$ l size of a code with distance $d$hef polynomials}
%right arrow
\newcommand{\rarrow}{\rightarrow}
%left arrow

\newcommand{\larrow}{\leftarrow}
%right arrow

\overfullrule=0pt
\def\setof#1{\lbrace #1 \rbrace}

\begin{abstract}

The first linear programming bound (\cite{mrrw}) is the best known asymptotic upper bound for binary codes, for a certain subrange of distances. Starting from the work of \cite{ft}, there are, by now, some arguably easier and more direct arguments for this bound. We show that this more recent line of argument runs into certain difficulties if one tries to go beyond this bound (say, towards the second linear programming bound of \cite{mrrw}).

\end{abstract}

\section{Introduction}

\noi A binary error-correcting code $C$ of length $n$ and minimal distance $d$ is a subset of the Hamming cube $\H$ in which the distance between any two distinct points is at least $d$. Let $A(n, d)$ be the maximal size of such a code.
In this note we are interested in the case in which the distance $d$ is linear in the length $n$ of the code, and we let $n$ go to infinity.
In this case $A(n, d)$ is known (see e.g., \cite{lev-chapter}) to grow exponentially in $n$, and we consider the quantity
\[
R(\delta) ~=~ \limsup_{n \rarrow \infty} \frac 1n \log_2 A\(n,\lfloor \delta n \rfloor\),
\]
also known as the {\it asymptotic maximal rate} of the code with relative distance $\delta$, for $0 \le \delta \le \frac12$.

\noi The best known upper bounds on $R(\delta)$ were obtained in \cite{mrrw} using the linear programming relaxation, constructed in \cite{dels}, of the combinatorial problem of bounding $A(n,d)$. At this point let us just note (more details will be provided in Section~\ref{sec:main}) that in \cite{mrrw} and in the follow-up result \cite{rodemich} two families of feasible solutions to the dual linear program of \cite{dels} for the Hamming cube  were constructed, leading to {\it the first and the second linear programming bounds} on the asymptotic rate function $R(\delta)$. (For brevity, we will refer to these bounds as the first and the second JPL bounds.) These constructions are not trivial, making use of the theory of orthogonal polynomials. The first JPL bound is easy to state and it will be stated below. The second bound is in general better, is obtained by a more complicated construction, and is harder to state explicitly. 

\noi Starting from the work of \cite{ft}, there are, by now,  some arguably easier and more direct arguments for the first JPL bound (see also \cite{ns,llc,ll}. Again, let us provide only a vague description (and moreover one applying with varying degree of accuracy) of the new line of argument in these results, making things more precise in the next section. In this line of argument, a feasible solution to the dual linear program of \cite{dels} for the Hamming cube is constructed in two steps. In the first step one finds a pair of auxiliary functions on $\H$, which have certain properties and which interact in a certain way. Given such a pair of functions, it is easy to construct a feasible solution to the dual linear program of \cite{dels}. Roughly speaking, the individual properties of these functions and of their interaction combine to supply the required properties of such a solution. Furthermore, and this is a key point, one can choose both of these functions and explain their interaction in a relatively simple manner. 

\noi The main observation of this note we is that there are certain difficulties in taking this line of argument beyond the first linear programming bound. Specifically, one cannot choose the auxiliary functions in the argument to be "simple". This statement will be made more precise in Proposition~\ref{pro:main}.

\subsection{Background, definitions, and notation}
\label{subsec:backg}

\noi We view the Hamming cube $\H$ as a metric space, with the Hamming distance between $x, y \in \H$ given by $|x - y| = |\{i: x_i \not = y_i\}|$. The Hamming weight of $x \in \H$ is $|x| = |\{i: x_i = 1\}|$. For $x, y \in \H$, we write $x+y$ for the modulo $2$ sum of $x$ and $y$. 
The {\it Hamming sphere} of radius $r$ centered at $x$ is the set $S(n,x,r) = \left\{y \in \H:~|x-y| = r\right\}$. The {\it Hamming ball} of radius $r$ centered at $x$ is the set $B(n,x,r) = \left\{y \in \H:~|x-y| \le r\right\}$. Clearly, for any $x \in \H$ and $0 \le r \le n$ we have $|S(n,x,r)| = {n \choose r}$ and $|B(n,x,r) = \sum_{k=0}^r {n \choose k}$.

\noi Note that the Hamming sphere $S(n,x,r)$ can be viewed as a (metric) subspace of $\H$. The Hamming cube and the Hamming sphere can also be viewed as graphs, with points of the space as vertices. Two vertices in the cube are adjacent if their Hamming distance is $1$. Two vertices in the sphere are adjacent if their Hamming distance is $2$. 

\noi We recall some basic notions in Fourier analysis on the boolean cube (see e.g., \cite{O'Donnel}, Section~1.5). For $S \in \H$, define the Walsh-Fourier character $W_S$ on $\H$ by setting $W_S(y) = (-1)^{\sum S_i y_i}$, for all $y \in \H$. The {\it weight} of the character $W_S$ is the Hamming weight of $S$.  The characters $\{W_S\}_{S \in \H}$ form an orthonormal basis in the space of real-valued functions on $\H$, under the inner product $\<f, g\> = \frac{1}{2^n} \sum_{x \in \H} f(x) g(x)$. The expansion $f = \sum_{S\in \H} \widehat{f}(S) W_S$ defines the Fourier transform $\widehat{f}$ of $f$. Explicitly, $\widehat{f}(S) = \<f,W_S\> = \frac{1}{2^n} \sum_{x \in \H} f(x) W_S(x)$.

\noi It will be convenient to denote by 
$\<\widehat{f},\widehat{g}\>_{\cal F} ~:=~ \sum_{S \in \H} \widehat{f}(S) \widehat{g}(S)$ 
the inner product of functions in the "Fourier domain". Note that this inner product is not normalized.

\noi The Parseval identity states that $\<f,g\> = \sum_{S\in \H} \widehat{f}(S) \widehat{g}(S) = \<\widehat{f},\widehat{g}\>_{\cal F}$.

\noi The {\it convolution} of $f$ and $g$ is defined by $(f \ast g)(x) ~=~ \frac{1}{2^n} \sum_{y \in \H} f(y) g(x+y)$.
The convolution transforms to dot product:  $\widehat{f\ast g} = \widehat{f} \cdot \widehat{g}$.

\noi {\it Symmetrization}. The symmetrizing operator ${\cal S}$  acts on functions on the boolean cube as follows: for $f:~\H \rarrow \R$, ${\cal S} f$ at a point $x$ is the average of the values of $f$ at all points with the same Hamming weight as $x$. Namely $\({\cal S} f\)(x) = \E_{|y|=|x|} f(y)$. It is easy to see that ${\cal S}$ commutes with Fourier transform, that is $\widehat{{\cal S} f} = {\cal S} \widehat{f}$.

\noi A function $f$ on the cube is {\it symmetric} if $f(x)$ depends only on the Hamming weight $|x|$. Note that $f$ is symmetric iff ${\cal S} f = f$. Furthermore, ${\cal S} f$ is symmetric for any $f$. By the properties of the symmetrizing operator ${\cal S}$, $f$ is symmetric if and only if $\widehat{f}$ is symmetric. 

\noi {\it Krawtchouk polynomials} (see e.g., \cite{mrrw}). For $0 \le i \le n$, let $F_s$ be the sum of all Walsh-Fourier characters of weight $s$, that is $F_s = \sum_{|S| = s} W_S$. It is easy to see that $F_s(x)$ depends only on the Hamming weight $|x|$ of $x$, and it can be viewed as a univariate function on the integer points $0,...,n$, given by the restriction to $\{0,...,n\}$ of the univariate polynomial $K_s = \sum_{k=0}^s (-1)^k {x \choose k} {{n-x} \choose {s-k}}$ of degree $s$, that is, $F_s(x) = K_s(|x|)$. The polynomial $K_s$ is the $s^{th}$ {\it Krawtchouk polynomial}. Abusing notation, we will also call $F_s$ the $s^{th}$ Krawtchouk polynomial, and write $K_s$ for $F_s$ when the context is clear.  

\noi We record one property of the Krawtchouk polynomials, which we will use below: For any $0 \le i, j \le n$ holds ${n \choose j} K_i(j) = {n \choose i} K_j(j)$. This is usually referred to as {\it reciprocity} of Krawtchouk polynomials.

\section{Linear programming bounds for binary codes} 
\label{sec:main}

\noi We provide a few more details on the approach of \cite{mrrw, rodemich}. Then we describe the more recent line of argument for the first JPL bound. Following that, we present our main claim. 

\noi Let us start with recalling that the linear programming approach of \cite{dels} applies in a more general setting of finite distance transitive metric spaces, that is spaces in which for any two pairs of equal-distanced points, there is an isometry taking the first pair onto the second one. The two spaces relevant for us here are the Hamming cube and the Hamming sphere. 

\noi For the Hamming cube, the linear program of \cite{dels} has the following simple description in the language of Fourier analysis on $\H$ (to the best of our knowledge, this is due to \cite{kl-personal}): 

\dfn
\label{dfn:Desarte-Fourier}

A non-zero function $\Lambda$ on $\H$ is a {\it feasible solution} to the (dual) linear program of \cite{dels} if it has the following properties. 

\begin{enumerate}

\item $\Lambda \ge 0$ 

\item $\widehat{\Lambda}(x) \le 0$ for $|x| \ge d$.

\item $\Lambda$ is {\it symmetric}, that is $\Lambda(x)$ depends only on $|x|$.

\end{enumerate}

\edfn

\thm (\cite{dels})
\label{thm:Delsarte}
Let $C$ be a binary error-correcting code $C$ of length $n$ and minimal distance $d$. Let $\Lambda$ be a function satisfying the conditions of Definition~\ref{dfn:Desarte-Fourier}. Then
\[
|C| ~\le~ \frac{2^n \widehat{\Lambda}(0)}{\Lambda(0)}.
\]
\ethm

\noi Let us make two comments about these statements. 

\myblt Theorem~\ref{thm:Delsarte} is shown in \cite{dels} in the more general setting of finite distance transitive spaces (and even more generally, association schemes). In the special case of the Hamming cube the theorem has an easy proof via Fourier analysis (\cite{kl-personal, ll}). 

\myblt Condition 3 in Definition~\ref{dfn:Desarte-Fourier} is 'inherited' from the association scheme framework. As observed in \cite{kl-personal, ll} it is not necessary. With that, it can be assumed to hold without loss of generality, since a function $\Lambda$ which satisfies the first two conditions of the definition can be {\it symmetrized} (see Section~\ref{subsec:backg}), while preserving $\Lambda(0)$ abd $\widehat{\Lambda}(0)$.

\noi In \cite{mrrw} and \cite{rodemich} two families of functions satisfying the conditions of Definition~\ref{dfn:Desarte-Fourier} were constructed, leading to two bounds on the asymptotic rate function $R(\delta)$. These constructions employ the theory of orthogonal polynomials, following \cite{dels}, where the problem of finding the optimal feasible solution for the dual linear program was formulated as an extremal problem for orthogonal polynomials. In these terms, the first family of functions, constructed in \cite{mrrw}, presents a certain (not necessarily optimal) solution to the extremal problem posed in \cite{dels}, in the special case of the Krawtchouk polynomials (a family of orthogonal polynomials associated with the Hamming cube, see Section~\ref{subsec:backg}). This construction leads to the {\it first JPL bound} 
\beqn
\label{ineq:JPL1}
R(\delta) ~\le~ H\(\frac12 - \sqrt{\delta(1-\delta)}\).
\eeqn
Here $H(x) = x \log_2\(\frac 1x\) + (1-x) \log_2\(\frac{1}{1-x}\)$ is the binary entropy function. 

\noi The {\it second JPL bound} in \cite{mrrw} is obtained in several steps. First, one bounds the cardinality of metric ball packings in a different metric space, the Hamming sphere $S(n,r) := S(n,0,r)$, by constructing a solution to the extremal problem of \cite{dels} for a different family of orthogonal polynomials, the {\it Hahn polynomials}, associated with this metric space. Then one connects, via the Bassalygo-Elias inequality, the bounds on metric ball packings in $S(n,r)$ and in $\H$, using the fact that $S(n,r)$ is isometrically embedded in $\H$. Finally, the second JPL bound is obtained by optimizing over the radius $r$. This bound is more complicated, and we do not state it explicitly. For our purposes it suffices to note that it coincides with the first linear programming bound for $\delta \ge 0.27...$, and it is better for smaller values of $\delta$. 

\noi Note that this procedure does not construct a new family of functions satisfying the conditions of Definition~\ref{dfn:Desarte-Fourier}. This last step was accomplished in \cite{rodemich}, where it was shown how to "lift" a feasible solution to the dual linear program of \cite{dels} for the Hamming sphere to a feasible solution to the dual linear program for the Hamming cube. Let us observe that the obtained family of functions is relatively complicated to understand, and it can be argued that it preserves much of its extrinsic, 'spherical' origins. 

\noi An arguably simpler and more direct proof of the first linear programming bound for binary linear codes was presented in \cite{ft}. In follow-up work (\cite{ns, llc}) this line of argument was extended to general binary codes. Furthermore, it was observed that this extended argument in fact stays within the framework of the linear programming approach of \cite{dels} and can be interpreted as providing an alternative way of constructing feasible solutions to the dual linear program of \cite{dels} (see the discussion preceding and following Theorem~1.1 in \cite{llc}). We proceed with presenting a high-level view of this approach. 

\lem
\label{lem:DH}

Let $f \ge 0$ and $g$ be two functions on $\H$, and let $\tau > \rho$ be two real numbers such that the following holds:

\begin{enumerate}

\item $g \ast f \ge \tau f$

\item $\widehat{g}(x) \le \rho$ if $|x| \ge d$.

\end{enumerate}

Let $\Lambda = g \ast f \ast f - \rho \cdot (f \ast f)$. Then $\Lambda$ satisfies the conditions of Definition~\ref{dfn:Desarte-Fourier}.

\elem

\cor 
\label{cor:alt LP bound}

Let $C$ be a code of length $n$ and minimal distance $d$. Then
\[
|C| ~\le~ \frac{\widehat{g}(0) -\rho}{\tau - \rho} \cdot  \frac{2^n \widehat{f}^2(0)}{\<f,f\>}.
\]

\ecor

\noi Let us make some comments about these results.

\myblt The claim of Corollary~\ref{cor:alt LP bound} follows immediately from Lemma~\ref{lem:DH} and Theorem~\ref{thm:Delsarte}.

\myblt Lemma~\ref{lem:DH} is a combination of two observations, Lemma~3.1 in \cite{ns} and an observation following Theorem~1.1 in \cite{llc}. Both observations were stated in a more restricted way, with $g$ taken to be proportional to the characteristic function of the Hamming sphere $S(n,1)$, but they extend essentially verbatim. 

\myblt It seems expedient to choose the function $g$ supported in a small Hamming ball around zero. In this case the first condition of Lemma~\ref{lem:DH} becomes a local one, and seems easier to control. In particular, if $g = 2^n \cdot 1_{S(n,1)}$, the function $g \ast f$ sums at each point of $\H$ the values of $f$ at its neighbors. For this choice of $g$ the first condition of Lemma~\ref{lem:DH} has a natural interpretation in terms of eigenvalues of subsets of $\H$ (\cite{ft,sam-log-sob}).  

\myblt Since we require $f$ to be nonnegative, it also seems reasonable to choose $g$ to be nonnegative, to facilitatethe first condition of Lemma~\ref{lem:DH}. 

\myblt Similarly, it seems reasonable to choose $g$ to be symmetric, to facilitate control of the second condition of Lemma~\ref{lem:DH}.

\noi We proceed with considering some examples. The third of these examples is just a restatement of Theorem~\ref{thm:Delsarte}. The first two examples recover the first JPL bound, employing the following useful observation. By an application of the Cauchy-Schwarz inequality, the bound in Corollary~\ref{cor:alt LP bound} can be replaced by $\frac{\widehat{g}(0) -\rho}{\tau - \rho} \cdot | \mathrm{supp}(f)|$.

\begin{itemize}

\item In \cite{ns} we choose $g = 2^n \cdot 1_{S(n,1)}$, we choose $f$ to be the maximal eigenfunction of the Hamming ball of radius $d' \approx \frac n2 - \sqrt{d(n-d)}$. We also choose $\rho = n - 2d$ and $\tau = n - 2d + 1$. The choice of $d'$ ensures that the maximal eigenvalue of the ball is at least $\tau$, which implies the first condition of Lemma~\ref{lem:DH}. The second condition of the lemma is easy to verify.
    
\item In \cite{ll} we choose $m = o(n)$, and we choose $g$ to be the $m$-wise convolution of $g_0$ with itself, where $g_0$ is the choice of $g$ in the preceding example. We choose $f = 1_{S(n,d')} + 1_{S(n,d'-1)}$, where $d' \approx \frac n2 - \sqrt{d(n-d)}$. We choose $\rho = (n - 2d)^m$ and $\tau = (n - 2d)^m + 1$. The first condition of Lemma~\ref{lem:DH} is verified by counting random walks in $\H$. The second condition of the lemma is easy to verify.
    
\item Let $\Lambda_0$ be a feasible solution to the dual linear program of \cite{dels}. Choose $g = \Lambda_0$, $f = 2^n \cdot 1_{\{0\}}$, $\tau = \frac{\Lambda_0(0)}{2^n}$, $\rho = 0$. It is easy to verify that the conditions of Lemma~\ref{lem:DH} are satisfied. Furthermore, let $\Lambda$ be the function constructed in the lemma. Then $\Lambda = \Lambda_0$.
    
\end{itemize}
    
\noi {\it Discussion}.

\myblt The third example shows that any feasible solution to the dual linear program of \cite{dels} can be obtained following the approach suggested in Lemma~\ref{lem:DH}. With that, this example obviously provides no new information. The usefulness of the first and the second examples are in the fact that both $f$ and $g$ are relatively simple objects. In the first example, $g$ is very simple, and in a way the "burden of proof" is delegated to $f$, which is somewhat more complicated. In the second example $f$ is simplified, at the cost of a certain increase in the complexity of $g$.

\myblt In all the examples above $g$ is nonnegative and symmetric. Furthermore, in the first two examples it carries a combinatorial meaning of counting walks (of different lengths) in $\H$, which is consistent with these properties. 

\myblt In all the examples $f$ is symmetric as well. While, as discussed above, choosing $g$ symmetric seems to be useful, making this choice for $f$ does not seem to be obviously helpful towards the goal of constructing both $f$ and $g$ as simple objects. And if $f$ is not symmetric, it might be advantageous to choose $g$ not to be symmetric as well. On the other hand, if $f$ is symmetric, it is easy to see that we can symmetrize $g$ as well, without loss of generality.

\subsection{Beyond the first linear programming bound}

\noi Recall that the first linear programming bound $R(\delta) \le H\(\frac12 - \sqrt{\delta(1-\delta)}\)$ is worse than the second linear programming bound for $0 < \delta < 0.27...$. Moreover, as $\delta$ gets smaller, it becomes progressively worse compared to other known bounds. In particular, for small $\delta$ it is worse than the packing bound $R(\delta) \le 1 - H\(\frac{\delta}{2}\)$ (\cite{Hamming}). This, and the relative complexity of the second linear programming bound, provides motivation to look for an easier argument for this bound, or at least for a bound which improves on the first JPL bound, along the lines of Lemma~\ref{lem:DH}. In this note we argue that this is not an easy task, if one makes a "natural" choice for the function $g$, as discussed above.

\noi One way to recover the second JPL bound is by proving an analog of the first JPL bound in the Hamming sphere, using an argument similar to that in Lemma~\ref{lem:DH},\footnote{This argument, with the choice of $f$ and $g$ as in the first example above, can be extended to other finite distance-transitive spaces (\cite{sam-formal ft - manuscript}) and also to some infinite symmetric spaces, such as the Euclidean sphere (\cite{noa-thesis}).} and then deriving the bound for the cube via the Bassalygo-Elias inequality, as described above. For our purposes, this approach is not very useful, since it is "extrinsic" to the cube, and hence does not seem to provide a simpler "intrinsic" construction for a family of functions satisfying the conditions of Definition~\ref{dfn:Desarte-Fourier} and improving on the first JPL bound. Such a construction might also have the potential dividend of suggesting a way to go beyond the analogs of the first JPL bound in other distance transitive spaces (\cite{Cohn}). 

\noi Proceeding with the attempt to improve the first JPL bound along the lines of Lemma~\ref{lem:DH}, we first note that choosing the "simplest possible" $g$, 
that is $g = 2^n \cdot 1_{S(n,1)}$, as in the first example above, will not work. As shown in \cite{sam-log-sob}, one can not go beyond the first JPL bound for this choice of $g$. With that, it might seem plausible that even the next simplest construction of $g$, taking it to be a nonnegative combination of the characteristic functions of Hamming spheres of radii $1$ and $2$ might suffice. In fact, one could reason as follows. Recall that for $g = 2^n \cdot 1_{S(n,1)}$, $g$ acts on $f$ via convolution, by summing at each point of $\H$ the values of $f$ at its neighbors. A corresponding operator on the Hamming sphere would sum at each point the values of a function at its neighbors in the sphere. Viewing the sphere as embedded in the cube, in a possible attempt to make the argument of \cite{rodemich} intrinsic to the cube, this "lifts" to summing  the values of a function at points at distance $2$ from it in the cube. 

\noi In the next claim we show this intuition to be false in a rather strong sense, provided we make the natural assumptions that $g$ is nonnegative and symmetric. 

\pro 
\label{pro:main}

Let $f$, $g$, $\tau$ and $\rho$ be as in Lemma~\ref{lem:DH}, with additional stipulations that $g$ is nonnegative and symmetric. Define $\delta = \frac dn$. Let $\e > 0$ be an absolute constant, and assume that 
\[
\frac 1n \log_2 \frac{\widehat{g}(0) -\rho}{\tau - \rho} \cdot  \frac{2^n \widehat{f}^2(0)}{\<f,f\>} ~\le~ H\(\frac12 - \sqrt{\delta(1-\delta)}\) - \e.
\]

Then, there exists a constant $\e'$ depending only on $\delta$ and $\e$, such that the support of $g$ is not contained in the Hamming ball of radius $\e' n$ around $0$. 
\epro

\noi {\it Discussion}

\myblt Recalling the first JPL bound (\ref{ineq:JPL1}), the assumption of the proposition is that for this choice of $f$ and $g$, the bound in Corollary~\ref{cor:alt LP bound} is better than the first JPL bound.

\myblt The conclusion of the proposition is that to beat the first JPL bound one has allow the action of $g$ on $f$ to be "complex" in one of the following ways. Either the support of $g$ is large and hence its action is not local (and hence the first condition of Lemma~\ref{lem:DH} is hard to enforce); or $g$ is allowed to be negative, and hence the first condition of Lemma~\ref{lem:DH} will involve some nontrivial cancellations; or both $f$ and $g$ are not symmetric, and hence their individual properties and their interaction are harder to enforce and to validate.

\section{Proofs}

\subsubsection*{Proofs of Lemma~\ref{lem:DH} and Corollary~\ref{cor:alt LP bound}}

\noi Both of these proofs are essentially given in \cite{ns,llc}. They are very short and we restate them for completeness. 

\noi {\bf Proof of Lemma~\ref{lem:DH}}. Since $f \ge 0$, we have $g \ast f \ast f \ge \tau \cdot f \ast f$, by the first condition of the lemma. Hence  $\Lambda = g \ast f \ast f - \rho \cdot (f \ast f) \ge (\tau - \rho) \cdot (f \ast f) \ge 0$. Next, since the convolution transforms to dot product, we have for $|x| \ge d$, $\widehat{\Lambda}(x) = \(\widehat{g}(x) - \rho\) \cdot \widehat{f}^2(x) \le 0$, by the second condition of the lemma. \eprf

\noi \noi {\bf Proof of Corollary~\ref{cor:alt LP bound}}. Note that $\Lambda(0) = \<g \ast f, f\> - \rho \<f,f\> \ge (\tau - \rho) \cdot \<f,f\>$, and that $\widehat{\Lambda}(0) = \(\widehat{g}(0) - \rho\) \cdot \widehat{f}^2(0)$. The claim of the corollary now follows from Theorem~\ref{thm:Delsarte}. \eprf

\subsubsection*{Proof of Proposition~\ref{pro:main}}

\noi We may and will assume that $n$ is sufficiently large (otherwise the claim of the proposition is satisfied trivially). Assume that the conditions of the proposition are satisfied and that for some $0 \le r \le n$, the support of $g$ is contained in the Hamming ball of radius $r$ around $0$. We proceed in two steps. In the first step we show that this implies the existence of a polynomial $P$ in one real variable of degree at most $r$, of a probability distribution vector $\(\l_0,\ldots,\l_n\)$ on $0,\ldots,n$, and of a constant $\delta < \beta < \frac12$ with the following properties (recall that $d = \delta n$):

\begin{enumerate}

\item $\sum_{\beta n \le i \le (1-\beta) n} \l_i \ge 1 - 2^{-\e_1 n}$, where $\e_1$ is an absolute constant depending on $\beta, \delta, \e$.

\item $P$ is a nonnegative combination of the Krawtchouk polynomials $K_0,\ldots,K_r$.

\item $\sum_{i=0}^n \l_i P(i) \ge P(d)$.

\end{enumerate}

\noi In the second step we will argue that this implies $r \ge \e' n$, where $\e'$ is an absolute constant depending on $\beta, \delta, \e_1$.

\noi {\it The first step}. We start with observing that the condition $g \ast f \ge \tau f$ implies $\widehat{f}(0) \widehat{g}(0) = \E(g \ast f) \ge \tau \E f = \tau \widehat{f}(0)$. Since $\widehat{f}(0) > 0$, this implies $\widehat{g}(0) \ge \tau$, and hence $\frac{\widehat{g}(0) -\rho}{\tau - \rho} \ge 1$. This means that the assumptions of the proposition imply that for some $\frac12 > \alpha > \delta$ we have
\[
\frac 1n \log_2 \frac{2^n \widehat{f}^2(0)}{\<f,f\>} ~\le~ \(H\(\frac12 - \sqrt{\delta(1-\delta)}\) - \e\) \cdot n ~=~ H\(\frac12 - \sqrt{\alpha(1-\alpha)}\) \cdot n,
\] 
Let $\l_i = \frac{\sum_{|x| = i} \widehat{f}^2(x)}{\<f,f\>}$, for $0 \le i \le n$. By the Parseval identity, $\(\lambda_0,\ldots,\l_n\)$ is a probability vector. Let $\delta < \beta < \alpha$. By Theorem~9 in \cite{PS}, we have $\sum_{\beta n \le i \le (1-\beta) n} \l_i \ge 1 - 2^{-\e_1 n}$, where $\e_1$ is an absolute constant depending on $\alpha$ and $\beta$. This proves the first of the above stated properties. 

\noi Next, recall that $g$ and hence $\widehat{g}$ are symmetric functions. For $0 \le i \le n$, write $a_i$ for $g(x)$ with $|x| = i$ and $b_i$ for $\widehat{g}(y)$ with $|y| = i$. Recall that by assumption $a_i \ge 0$ and $a_i = 0$ for $r+1 \le i \le n$. Define a function $P$ on the integer points $0,\ldots,n$ by setting $P(i) = b_i$. It is an easy and standard corollary of the properties of the Krawtchouk polynomials $K_0,\ldots,K_n$, and in particular of their reciprocity, that $P = \sum_{i=0}^r a_i K_i$ (and in particular, $P$ is a polynomial of degree at most $r$). This proves the second property. 

\noi To prove the third property, recall that by assumption $P(i) \le \rho$ for $d \le i \le n$, and in particular $P(d) \le \rho$. Hence, using the assumptions of the proposition, in particular the fact that $\widehat{g}$ is symmetric in the penultimate step,
\[
P(d) \le \rho < \tau \le \frac{\<g \ast f, f\>}{\<f,f\>} ~=~ \sum_{x \in \H} \widehat{g}(x) \frac{\widehat{f}^2(x)}{\<f,f\>} ~=~ \sum_{i=0}^n b_i \cdot \frac{\sum_{|x| = i} \widehat{f}^2(x)}{\<f,f\>} ~=~ \sum_{i=0}^n \l_i P(i). 
\]
   
\noi {\it The second step}. We require the following properties of the Krawtchouk polynomials (see e.g., \cite{lev-chapter}). Let $0 \le m \le n$. The Krawtchouk polynomial $K_m$ has positive value ${n \choose m}$ at $0$. This is also the maximal value of $K_m$ in the interval $[0,n]$. Furthermore, $K_m$ has $m$ distinct real roots $x_1 < x_2 < \ldots < x_m$ in the interval $(0,n)$. These roots are symmetric around $\frac n2$ and $x_1 \ge \frac n2 - \sqrt{m(n-m+2)}$. Assume now that $m$ is sufficiently small, so that $x_1 - 1 > \beta n$. Since $d < \beta n$, we deduce that $K_m(d)$ is positive and 
\[
\frac{K_m(d)}{K_m(0)} ~\ge~ \(\frac{x_1 - d}{x_1}\)^m ~\ge~ \(\frac{\beta n- d}{n/2}\)^m ~=~ (2\beta - 2\delta)^m.
\] 

\noi Similarly, for any $\beta n \le i \le x_1-1$, we have that $K_m(i)$ is positive and 
\[
\frac{K_m(i)}{K_m(d)} ~\le~ \frac{x_1 - i}{x_1 - d} ~\le~ \frac{x_1 - \beta n}{x_1 - \delta n} ~\le~ \frac{n/2 - \beta n}{\beta n - \delta n} ~=~ \frac{1 - 2\beta}{2 \beta - 2\delta}. 
\]

\noi By the symmetry of the roots $x_1 < x_2 < \ldots < x_m$ around $\frac n2$, the same is true for any $x_m +1 \le i \le (1-\beta) n$, replacing $K_m(i)$ by its absolute value. Finally, for any $x_1 \le i \le x_m$ we have $|K_m(i)| \le K_m\(x_1 - 1\) \le \frac{1 - 2\beta}{2\beta - 2\delta} \cdot P(d)$.

\noi Assume now that $r$ is sufficiently small, so that $x_1(r) - 1 > \beta n$, where $x_1(r)$ is the first root of $K_r$. Recall that the sequence $x_1(m)$ is decreasing in $m$ (\cite{lev-chapter}). From the preceding discussion, the inequalities above hold for all $0 \le m \le r$. Since $P$ is a nonnegative combination of $K_0,...,K_r$ they also hold for $P$ (replacing $m$ with $r$). It follows that 
\[
P(d) ~\le~ \sum_{i=0}^n \l_i P(i) ~=~  \sum_{i = 0}^{\beta n - 1} \l_i P(i) + \sum_{\beta n \le i \le (1-\beta) n} \l_i P(i) + \sum_{i = (1-\beta) n+1}^n \l_i P(i) ~\le~ 
\]
\[
\(1 - \sum_{\beta n \le i \le (1-\beta) n} \l_i\) \cdot P(0) + \max_{\beta n \le i \le (1-\beta) n} |P(i)| ~\le~ \(2^{-\e_1 n} \cdot \(\frac{1}{2\beta - 2\delta}\)^r + \frac{1 - 2\beta}{2\beta - 2\delta}\) \cdot P(d).
\]

\noi It is easy to see that this can hold only if $r$ is linearly large in $n$. This completes the proof of the proposition. \eprf

\end{document}